\documentclass[preprint,aps]{revtex4}

\usepackage{graphicx}

\begin{document}

\title{Fermi Surface and Band Structure of (Ca,La)FeAs$_2$ Superconductor from Angle-Resolved Photoemission Spectroscopy}
\author{Xu Liu$^{1,\sharp}$, Defa Liu$^{1,\sharp}$, Lin Zhao$^{1,\sharp,*}$, Qi Guo$^{1,\sharp}$, Qingge Mu$^{1}$, Dongyun Chen$^{1}$, Bing Shen$^{1}$, Hemian Yi$^{1}$, Jianwei Huang$^{1}$, Junfeng He$^{1}$, Yingying Peng$^{1}$, Yan Liu$^{1}$, Shaolong He$^{1}$, Guodong  Liu$^{1}$,  Xiaoli Dong$^{1}$, Jun Zhang$^{1}$, Chuangtian Chen$^{2}$, Zuyan Xu$^{2}$,  Zhi-an Ren$^{1,*}$, and X. J. Zhou$^{1,*}$
}

\affiliation{
\\$^{1}$National Laboratory for Superconductivity, Beijing National Laboratory for Condensed
Matter Physics, Institute of Physics, Chinese Academy of Sciences,
Beijing 100190, China
\\$^{2}$Technical Institute of Physics and Chemistry, Chinese Academy of Sciences, Beijing 100190, China
}
\date{November 23, 2013}
%
% The abstract goes here
%

\begin{abstract}
The (Ca,R)FeAs$_2$ (R=La,Pr and etc.) superconductors with a signature of superconductivity transition above 40 K possess a new kind of block layers that consist of zig-zag As chains. In this paper, we report the electronic structure of the new (Ca,La)FeAs$_2$ superconductor investigated by both band structure calculations and high resolution angle-resolved photoemission spectroscopy measurements. Band structure calculations indicate that there are four hole-like bands around the zone center $\Gamma$(0,0) and two electron-like bands near the zone corner  M($\pi$,$\pi$) in CaFeAs$_2$. In our angle-resolved photoemission measurements on (Ca$_{0.9}$La$_{0.1}$)FeAs$_2$, we have observed three hole-like bands around the $\Gamma$ point and one electron-like Fermi surface near the M($\pi$,$\pi$) point. These results provide important information to compare and contrast with the electronic structure of other iron-based compounds in understanding the superconductivity mechanism in the iron-based superconductors.
\end{abstract}

\pacs{74.70.-b, 74.25.Jb, 79.60.-i, 71.20.-b}

\maketitle

Since the first report of superconductivity in LaFeAsO$_{1-x}$F$_x$\cite{Kamihara}, several families of the iron-based superconductors have been discovered, including  ReFeAs(O,F) (so-called 1111 system where Re represents rare earth elements like La,Ce,Pr,Nd,Sm and etc.)\cite{Kamihara,XHChen43K,GFChenCe,ZARenNd,ZARenPr,ZARenSm}, (A,K)Fe$_2$As$_2$ (122 system where A represents alkaline earth elements like Ba, Sr and Ca)\cite{RotterSC,Sasmal}, CFeAs (111 system where C represents alkali metal elements like Li and Na)\cite{WangXC,JHTapp}, and FeSe (11 system)\cite{FCHsu}. All these iron-based superconductors share the same FeAs or FeSe layers which are believed to be essential for the occurrence of superconductivity in these compounds. Different families of the iron-based superconductors form because of the presence of various block layers that separate FeAs or FeSe layers, and  various stacking sequences of the FeAs (FeSe) layers with these block layers.  The latest report of superconductivity with a transition temperature T$_c$ as high as 43 K in (Ca,R)FeAs$_2$(where R=La and Pr)\cite{NKatayama,HYakita,KKudo} is interesting because it has a new form of block layers in its structure.  The crystal structure of (Ca$_{1-x}$La$_x$)FeAs$_2$\cite{NKatayama} consists of alternately stacked FeAs layers and arsenic layers that consist of arsenic zigzag chains with a formal electron count of As$^{-}$\cite{HYakita,NKatayama}. It is important to first understand the electronic structure of this new class of iron-based superconductors.

In this paper, we report the electronic structure of the new (Ca,La)FeAs$_2$ superconductor investigated by both band structure calculations and first high resolution angle-resolved photoemission spectroscopy (ARPES) measurements. Band structure calculations indicate that there are four hole-like bands around the zone center $\Gamma$(0,0) and two electron-like bands near the zone corner  M($\pi$,$\pi$) in CaFeAs$_2$. In our angle-resolved photoemission measurements, we have observed three hole-like bands around the $\Gamma$ point and one electron-like Fermi surface near the M($\pi$,$\pi$) point. These results provide important information to compare and contrast with the electronic structure of the other iron-based compounds in understanding the superconductivity mechanism in the iron-based superconductors.

The (Ca$_{1-x}$La$_x$)FeAs$_2$ single crystals were grown using flux method similar to that reported in\cite{NKatayama}. The grown single crystals have a typical size up to 0.5 mm $\times$ 0.5 mm $\times$ 0.05 mm (The morphology of one single crystal is shown in the inset of Fig. 1b).  The single crystals of (Ca$_{1-x}$La$_x$)FeAs$_2$ (x=0.10, nominal composition) we grew were first characterized by X-ray diffraction (XRD) analysis. For this purpose, we selected two pieces of single crystals from the same grown ingot. Single crystal XRD analysis was performed on one piece and the obtained pattern is shown in Fig. 1b. All the observed peaks are sharp and can be indexed to (00$\emph{l}$) reflections of the  (Ca$_{1-x}$La$_x$)FeAs$_2$ (x=0.10) crystal, indicating high purity of the crystal surface. The other piece of the single crystal was ground into powder and the obtained powder X-ray diffraction pattern is shown in Fig. 1c. It shows that the main peaks can be indexed to (Ca$_{1-x}$La$_x$)FeAs$_2$ with a small amount of impurity phase from FeAs (marked by blank triangles) that is self-solvent during the crystal growth process.  In addition, a trace of CaFe$_2$As$_2$ phase is also detected (marked by solid triangle). From the single crystal and powder XRD analyzes of our samples, we believe the presence of the tiny amount of CaFe$_2$As$_2$ phase will not affect our angle-resolved photoemission measurements on the (Ca$_{1-x}$La$_x$)FeAs$_2$ phase. Fig. 1d shows the magnetic measurement result on the (Ca$_{1-x}$La$_x$)FeAs$_2$ (x=0.10) single crystal. The onset superconducting transition temperature is $\sim$ 27.5 K for x=0.10 that is comparable to the result reported before\cite{NKatayama}.

We carried out band structure calculations on CaFeAs$_2$ using Wien2K software package\cite{Wien2K}, as shown in Fig. 2. Since pure CaFeAs$_2$ is not synthesized yet, the crystal structure, lattice constants, and atomic positions of CaFeAs$_2$ are taken from La-doped CaFeAs$_2$ in \cite{NKatayama}.  The lattice constants used are a=3.9471 $\AA$, b=3.8724 $\AA$ and c=10.3210 $\AA$ taking a space group of P2$_{1}$\cite{NKatayama}.  For comparison, we also did calculations on BaFeAs$_2$ and the obtained band structure is in complete agreement with that reported in \cite{JHShim}. Our calculated Fermi surface of CaFeAs$_2$ (Fig. 2c) is consistent with that calculated before in \cite{NKatayama}. As seen from Fig. 2a, the calculated band structure near $\Gamma$(0,0) consists of four hole-like bands, corresponding to four hole-like Fermi surface sheets shown in Fig. 2c. This is different from that of previous iron-based compounds like BaFe$_2$As$_2$ where three hole-like bands are present near the $\Gamma$ point\cite{DJSingh}.  On the other hand, near the M($\pi$,$\pi$) point, two electron-like bands are present (Fig. 2a) giving rise to two electron-like Fermi surface sheets near the M point (Fig. 2c). This is similar to that in other iron-based compounds like in BaFe$_2$As$_2$\cite{DJSingh}. As in other iron-based superconductors, the electronic structure in CaFeAs$_2$ is also mainly dictated by the FeAs-layers. We note that except for the small Fermi surface  pocket and another inner Fermi surface sheet around $\Gamma$, the band structure of CaFeAs$_2$ is quite two-dimensional.

The angle-resolved photoemission measurements were carried out using our lab system equipped with Scienta R4000 electron energy analyzer\cite{GDLiu}. We use Helium I resonance line as the light source which gives a photon energy of h$\upsilon$=21.218 eV. The light on the sample is partially polarized with the
electric field vector mainly in the plane of the sample surface. The energy resolution was set at 20 meV and the angular resolution is $\sim$0.3 degree. The
Fermi level is referenced by measuring the Fermi edge of a clean polycrystalline gold that is electrically connected to the sample.   The crystal was cleaved {\it in situ} and measured in vacuum with a base pressure better than 6$\times$10$^{-11}$ Torr.

Figure 3 presents Fermi surface and constant energy contours at different binding energies of (Ca$_{0.9}$La$_{0.1}$)FeAs$_2$  measured at a temperature of 30 K. The corresponding band structures along several typical high symmetry cuts are shown in Fig. 4.  Around the $\Gamma$ point, two clear features can be identified. The first is a small strong-intensity patch confined near the $\Gamma$ point that is marked by red dashed line in Fig. 3a. The second obvious feature is the large Fermi surface that is nearly square-like. The spectral weight suppression along the $\Gamma$-M2(-$\pi$,$\pi$) direction of this square-shaped Fermi surface is likely due to photoemission matrix element effect. With increasing binding energy, the area of the square-shaped sheet grows indicating its hole-like nature that can be directly seen from the band structure measured along a momentum cut across the $\Gamma$ point in Fig. 4(d,g,j). Around the M2(-$\pi$,$\pi$) and M3(-$\pi$,-$\pi$) points at the corners of the first Brillouin zone, one Fermi surface sheet can be discerned. The corresponding band structure measurements (Fig. 4(e,h,k) and Fig. 4(f,i,l) for M2 and M3 cuts, respectively) indicate that the Fermi surface sheet is electron-like around the M points. The slight Fermi momentum difference seen between the M2 and M3 cuts (Figs. 4k and 4l) suggests that its shape is elliptical.

From the measured band structure around the $\Gamma$ point (Cut 1, Fig. 4(d,f,j)), we can identify three hole-like bands. The outer one with large Fermi momentum, denoted as $\gamma$ band, is most obvious. It has a Fermi momentum at 0.27 $\AA$$^{-1}$ and a Fermi velocity of 0.27 eV$\cdot$$\AA$; the corresponding effective mass of this $\gamma$ band is  7.6 m$_e$.  The second hole-like band around $\Gamma$ is denoted as $\beta$ in Fig. 4j. Its top nearly touches the Fermi level.  The third hole-like band around the $\Gamma$ point is weak but discernable, and is denoted as $\alpha$ band in Fig. 4j. The top of the $\alpha$ band also nearly touches the Fermi level.   Around the M points, we can see only electron-like band denoted as $\delta$ band (Fig. 4k and 4l). From the band structure near M2 (Fig. 4k), we obtain a Fermi momentum of this $\delta$ band as 0.22 $\AA$$^{-1}$, its Fermi velocity and corresponding effective mass are  0.55 eV$\cdot$$\AA$ and 3.1 m$_e$, respectively. From the band structure near M3 (Fig. 4l), the obtained Fermi momentum, Fermi velocity and effective mass are 0.13 $\AA$$^{-1}$,   0.74 eV$\cdot$$\AA$ and  1.4 m$_e$, respectively.

The measured electronic structure of (Ca$_{0.9}$La$_{0.1}$)FeAs$_2$ superconductor is reminiscent of that for the other iron-based superconductors like (Ba,K)Fe$_2$As$_2$\cite{DJSingh,LZhao1,HDing}. When compared with the band structure calculations (Fig. 2), the measured Fermi surface (Fig. 3) and band structure (Fig. 4) show an overall agreement but with some differences. Among the four hole-like Fermi surfaces expected from the band structure calculations around $\Gamma$ (Fig. 2a), three are observed in the measured band structure (Fig. 4j). The observed strong intensity patch near $\Gamma$ is likely due to the top of the $\alpha$ and $\beta$ bands, and more probably due to the bottom of the existence of an electron-like band at $\Gamma$, as seen from Fig. 2a. Its exact origin needs further investigation.  The observation of an electron-like Fermi surface sheet near M is consistent with the band structure calculations although one would expect to see two electron-like Fermi surface sheets near M.  Taking two electron-like Fermi surface sheets near M, the Fermi surface area of the electron-like sheets around M is larger than the hole-like sheet around $\Gamma$, consistent with the expectation that the sample is electron-doped.  As seen from Fig. 3a, the nesting between the electron-like Fermi surface near M and the hole-like Fermi surface sheets near $\Gamma$ looks poor for this superconductor with a T$_c$ of 27.5 K. This suggests that Fermi surface nesting may not play a major role in dictating the superconductivity mechanism in this new iron-based superconductor\cite{Nesting}.

In summary, we have carried out band structure calculations and high resolution angle-resolved photoemission measurements to study the electronic structure of  the new iron-based superconductor (Ca$_{0.9}$La$_{0.1}$)FeAs$_2$. We have revealed hole-like Fermi surface sheets near the $\Gamma$ point and electron-like Fermi surface sheet near the M points. The measured electronic structure of (Ca$_{0.9}$La$_{0.1}$)FeAs$_2$ shows agreement with band structure calculations and exhibits similarities with that of other iron-based superconductors. But there are unique features related with this new iron-based superconductors like the expected existence of four hole-like bands around the $\Gamma$ point, and the possible involvement of As states in the low-energy states near the Fermi level. These results will provide information to compare and contrast with the electronic structures of other iron-based compounds in sorting out the key factors to understand the superconductivity mechanism of the iron-based superconductors. Further improvement of the single crystal sample quality is needed in order to further study the electronic structure details and the superconducting gap by ARPES in (Ca,R)FeAs$_2$ superconductors.

We thank financial support from the NSFC (10734120) and the MOST of China (973 program No: 2011CB921703 and 2011CB605903).\\

$^{\sharp}$These people contribute equally to the present work.\

$^{*}$Corresponding author: XJZhou@iphy.ac.cn, renzhian@iphy.ac.cn, and  lzhao@iphy.ac.cn\\

\newpage

\begin{figure*}[floatfix]
\begin{center}
\includegraphics[width=1.0\columnwidth,angle=0]{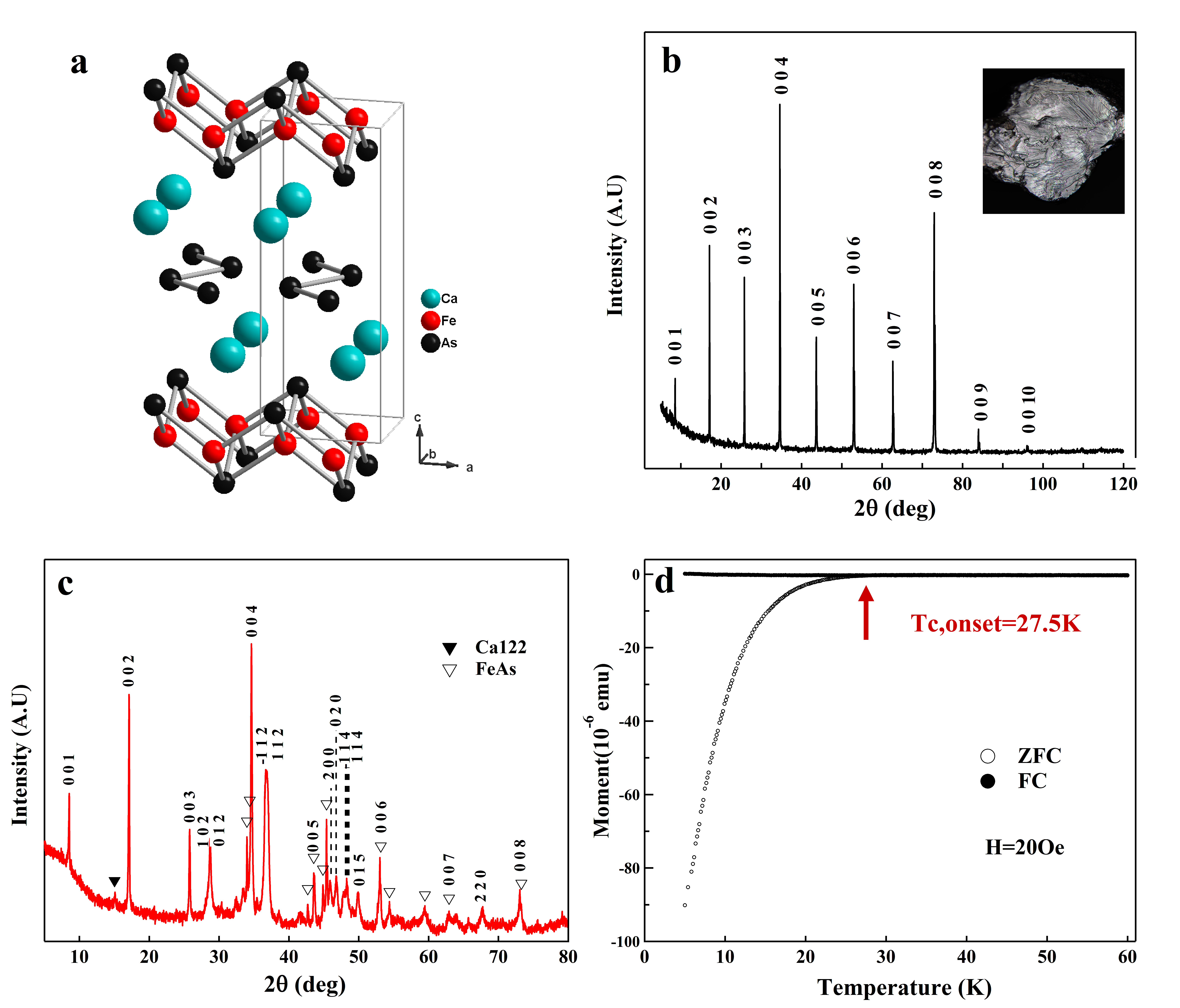}
\end{center}
\caption{Crystal structure, X-ray diffraction (XRD) pattern, and magnetic measurement result of
(Ca$_{1-x}$La$_x$)FeAs$_2$ (x=0.1). (a). Crystal structure of (Ca$_{1-x}$La$_x$)FeAs$_2$ (x=0.1) with a monoclinic structure\cite{NKatayama}. Grey dotted lines indicate the unit cell.  (b). XRD pattern of a plate-like crystal surface. The inset shows an optical image of the single crystal sample. (c). The powder XRD pattern. In this measurement, the powder was obtained by grinding a single crystal sample.  (d). Magnetic measurement of the single crystal sample that was used for the ARPES measurements presented in the work. The measurement was carried in both zero-field cooled (ZFC) and field cooled (FC) modes using a magnetic filed H=20 Oe. A superconducting transition temperature (T$_c$) of 27.5 K is observed.
}
\end{figure*}

\begin{figure}[tbp]
\begin{center}
\includegraphics[width=1.0\columnwidth,angle=0]{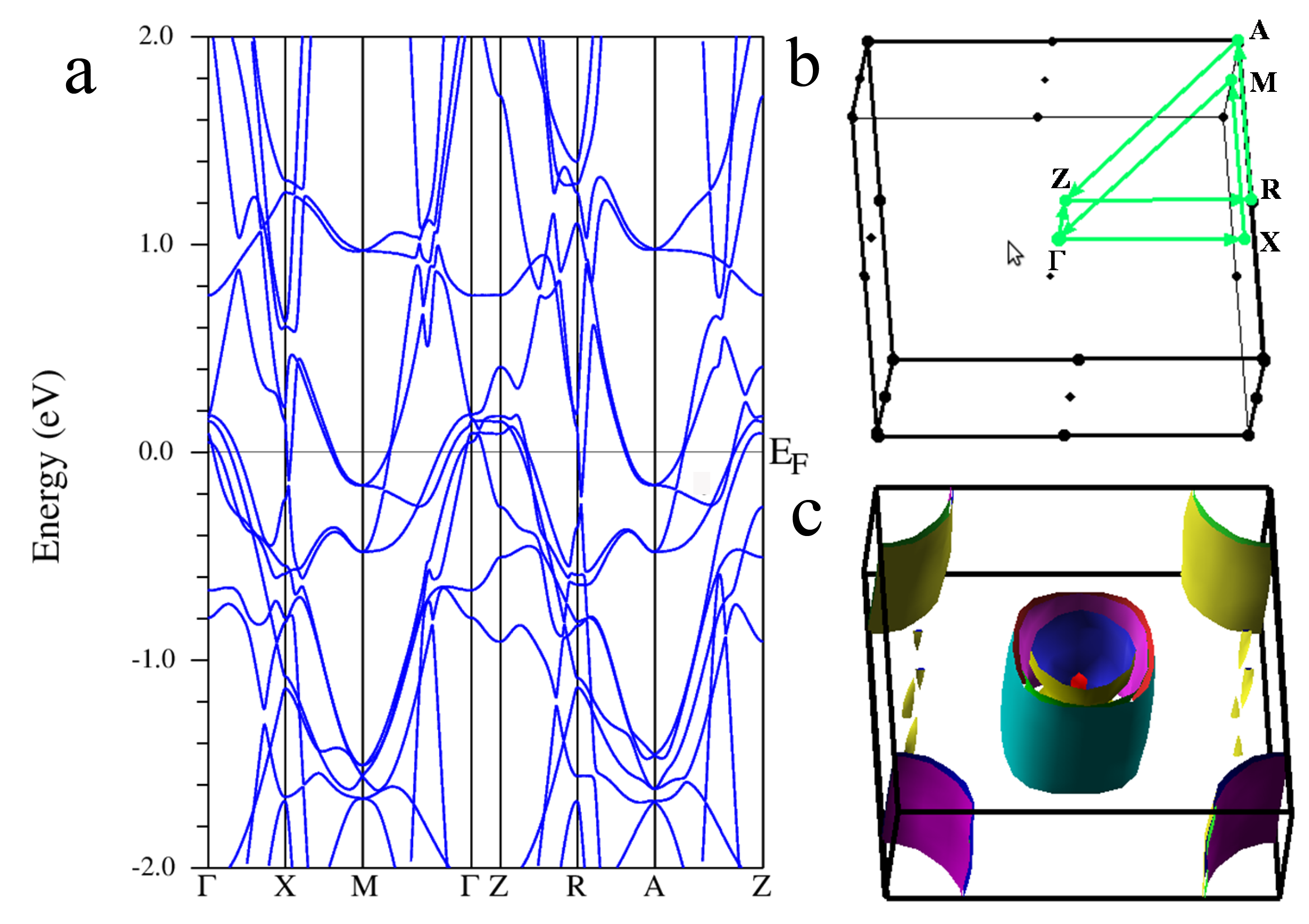}
\end{center}
\caption{Calculated electronic structure of CaFeAs$_2$. (a). The band structure along the high symmetry lines in the Brillouin zone.  (b). The first Brillouin zone and high symmetry points and lines. (c). The calculated Fermi surface.
}
\end{figure}

\begin{figure*}[floatfix]
\begin{center}
\includegraphics[width=1.0\columnwidth,angle=0]{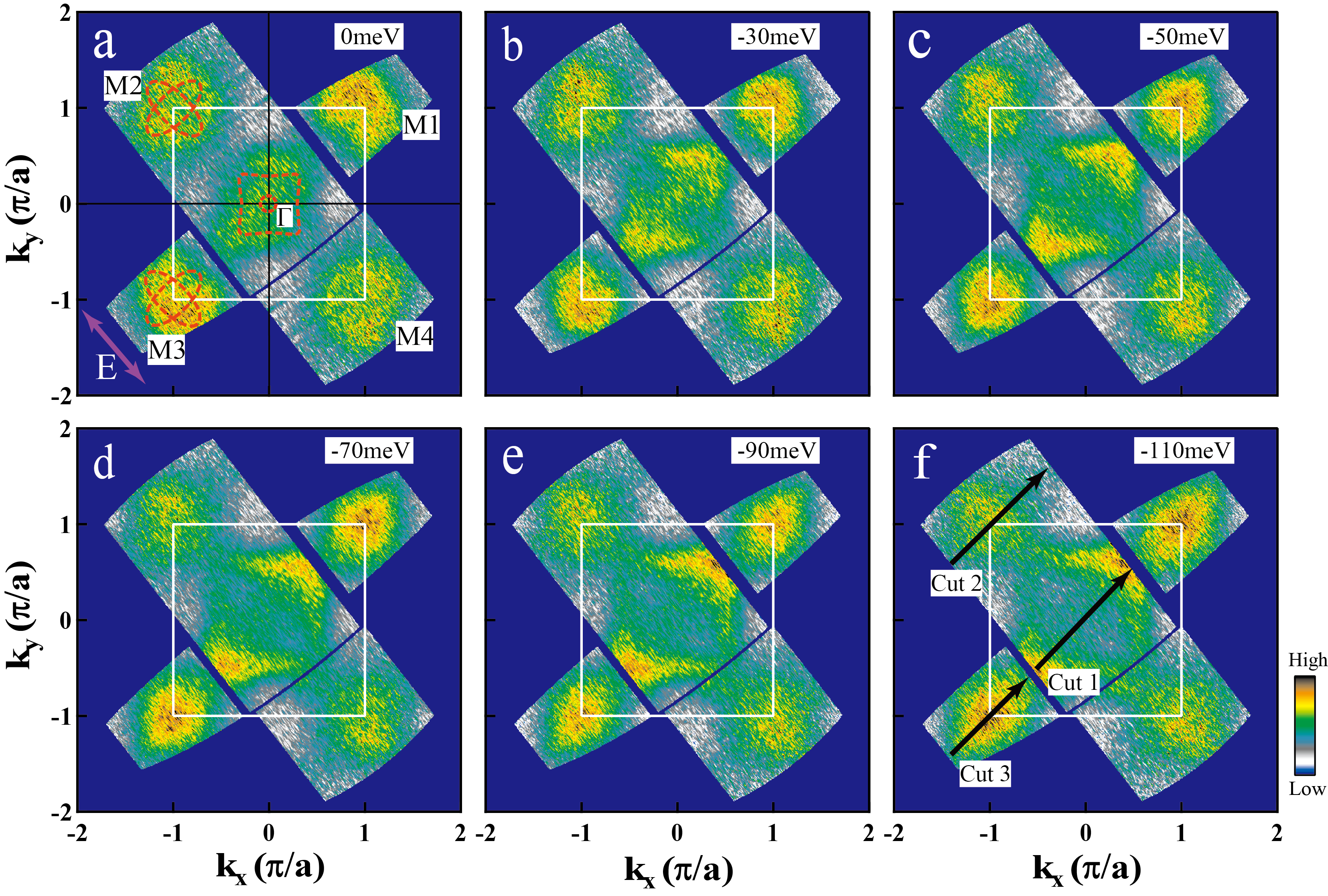}
\end{center}
\caption{Fermi surface and constant energy contours at different binding energies of (Ca$_{1-x}$La$_x$)FeAs$_2$ (x=0.1) measured at a temperature of 30 K.
(a-f). Spectral weight distribution integrated within a narrow energy window (20 meV) at various binding energies of 0 meV (a), 30 meV (b), 50 meV (c), 70 meV (d), 90 meV (e) and 110 meV (f),  as a function of k$_x$ and k$_y$. The electric field vector E of the partially polarized light source is  marked by pink line with arrows in (a).  The dashed red lines in (a) are guide to the eye for the Fermi surface topology. The three black lines in (f) represent the location of three momentum cuts where the measured band structures are presented in Fig. 4.  For convenience, the four equivalent M points are marked as M1, M2, M3 and M4, as shown in (a).
}
\end{figure*}

\begin{figure}[tbp]
\begin{center}
\includegraphics[width=0.85\columnwidth,angle=0]{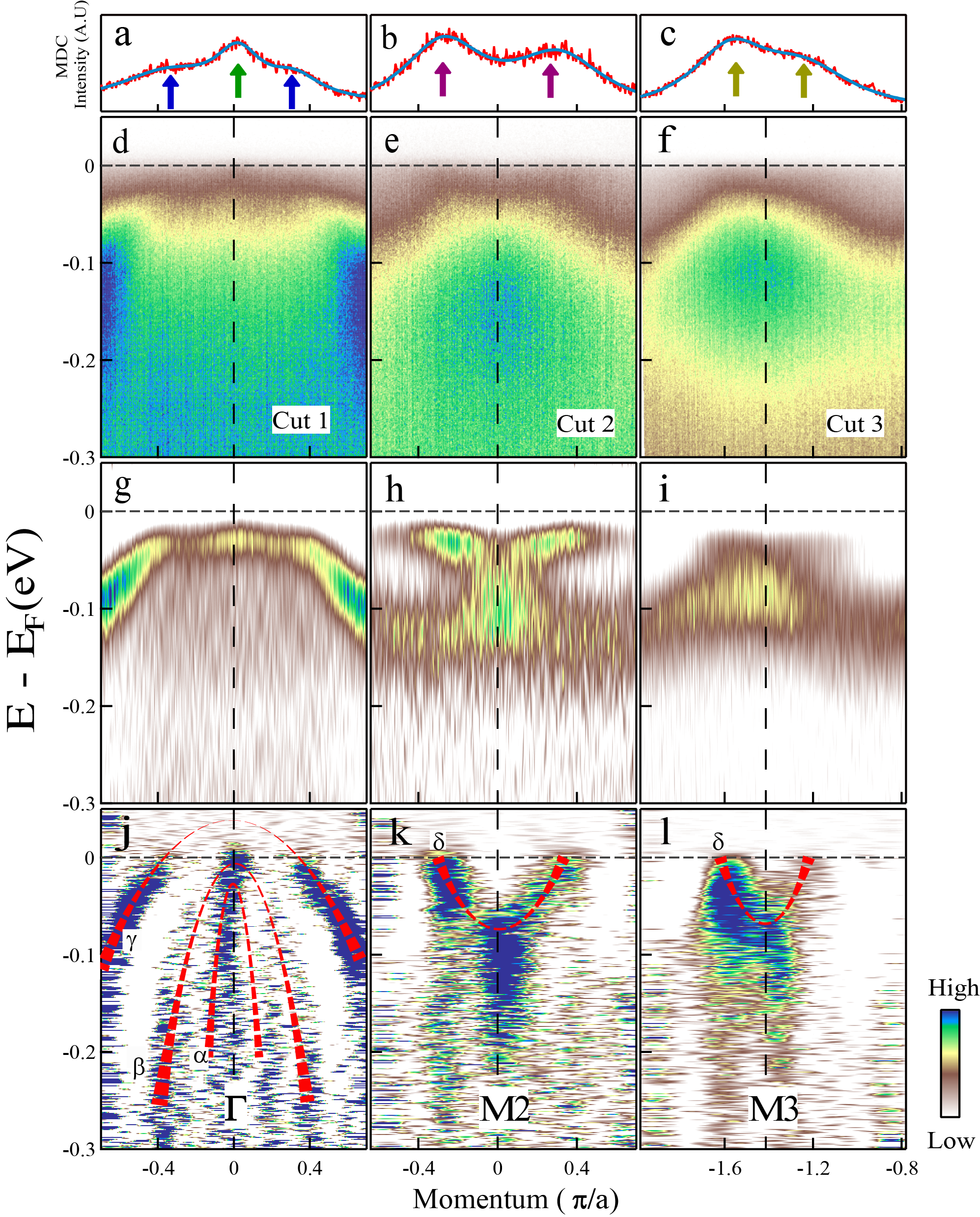}
\end{center}
\caption{Energy bands of (Ca$_{1-x}$La$_x$)FeAs$_2$ (x=0.1) measured along three high symmetry cuts. The location of  the three cuts are shown in Fig. 3f.  (d-f) show the original measured band structure along Cut 1, Cut 2, and Cut 3, respectively.  (g-i) shows the corresponding second derivative images with respect to the energy for the above three cuts;  (j-l) shows the corresponding second derivative images with respect to the momentum for the above three cuts.  The top panels(a-c) show the momentum distribution curves (MDCs) at the Fermi level for the photoemission image of Cut 1 to Cut 3. The arrows mark the location of the Fermi momenta.
}
\end{figure}

\end{document}